\documentstyle [prb,aps,epsf]{revtex}

\begin {document}

\draft

\title{
 Microscopic theory of the pseudogap and Peierls\\
  transition in quasi-one-dimensional materials
}

\author{Ross H. McKenzie\cite{email}}

\address{School of Physics, University of New
South Wales, Sydney, NSW 2052, Australia}

\date{Received 23 June 1995}
\maketitle
\begin{abstract}
The problem of deriving from microscopic theory
a Ginzburg-Landau free energy functional to
describe the Peierls or charge-density-wave transition
in quasi-one-dimensional materials is considered.
Particular attention is given to how  the thermal lattice motion
affects the electronic states.
 Near the transition temperature the thermal lattice
motion produces a
pseudogap in the density of states at the Fermi level.
Perturbation theory diverges and the traditional quasi-particle
or Fermi liquid picture breaks down.
The pseudogap causes a significant modification
of the coefficients in the Ginzburg-Landau functional
from their values in the rigid lattice approximation,
which neglects the effect of the thermal lattice motion.\\
\\
To appear in {\it Physical Review B} \\

\end{abstract}


\section{INTRODUCTION}
\subsection{Motivation}

A wide range of quasi-one-dimensional materials  undergo a
structural transition, known as the Peierls or charge-density-wave
(CDW) transition, as the temperature is lowered \cite{gru,con,gor,car}.
A periodic lattice distortion, with wave vector, $2k_F$, twice that of
the Fermi wavevector, develops along the chains.
Anomalies are seen in the electronic properties
due to the opening of an energy gap over the Fermi surface.

Over the past decade, due to the development of high-quality
samples and higher resolution experimental techniques,
new data has become available
which allows a quantitative comparison of experiment with theory.
The most widely studied material is the blue bronze, K$_{0.3}$MoO$_3$.
There is a well-defined three-dimensional transition at
$T_P=183$ K and careful measurements have been made of
thermodynamic anomalies \cite{bri} and CDW coherence lengths \cite{gir}
at the transition. The critical region, estimated from the
Ginzburg criterion \cite{gin} is only a few percent of
the transition temperature and so the transition should be described
by an anisotropic three-dimensional Ginzburg-Landau free energy
functional, except close to the transition temperature.
The challenge is to derive from a microscopic theory
the coefficients in the Ginzburg-Landau free energy
 so a quantitative comparison
can be made between theory and experiment.
Inspiration is provided by the case of superconductivity.
The superconducting transition is well described by
Ginzburg-Landau theory and the coefficients can be calculated
from BCS theory \cite{schr} and depend on microscopic
parameters such as the normal state density of states,
Debye frequency, and the electron-phonon coupling.
This program is so successful that one can even consider
refinements to BCS theory, such as strong coupling effects,
in order to get better agreement between experiment and
 theory \cite{carb}.
However, the problem of the CDW transition is more difficult
because of the large fluctuations due to the quasi-one dimensonality.

\subsection{Ginzburg-Landau theory}
The Peierls transition is described by an order parameter
which is proportional to the $2k_F$
lattice distortion along the chains.
The order parameter is complex if the lattice distortion
is incommensurate with the lattice. For a commensurate
lattice distortion (e.g., a half-filled band) the order
parameter is real.
I recently considered the general problem of Ginzburg-Landau
theory for a three-dimensional phase transition,
described by a complex order parameter,
in a system of weakly coupled chains \cite{mck2}.
The key results of that study are now summarized,
partly to put this paper in a broader context.

The Ginzburg-Landau free energy functional $F_1[\phi]$
for a {\it single} chain with a complex order
parameter $\phi(z)$, where $z$ is the co-ordinate along
the chain, is
\begin{equation}
F_1[\phi]=\int dz \left[
a \mid\phi\mid^2 + \ b  \mid\phi\mid^4 +
 \  c \mid {\partial \phi\over \partial z}\mid ^2 \right].
   \label{aa1}
\end{equation}
Near the single chain mean-field transition temperature
$T_0$ the second-order coefficient $a(T)$ can be written
\begin{equation}
a(T)= a^\prime \left( {T \over T_0} - 1 \right).
   \label{aa10}
\end{equation}
Due to fluctuations in the order parameter
this one-dimensional system cannot develop long-range order at
finite temperature \cite{lan,sca}.
To describe a finite-temperature
phase transition, consider a set of weakly interacting
chains.  If $\phi_i(z)$ is the order parameter on the $i$-th chain
the free energy functional for the system is
\begin{equation}
F[\phi_i(z)]=\sum_i F_1[\phi_i(z)] -
{J \over 4} \sum_{<i,j>} \int dz {\rm Re} [\phi_i(z)^*  \phi_j(z)]
   \label{ad1}
\end{equation}
where $J$ describes the interchain interactions
between nearest neighbours.
A mean-field treatment of this functional will only
give accurate results if the width of the three-dimensional
critical region is much smaller than $T_0$. This requires
that the width of the one-dimensional critical region $\Delta t_{1D}
\equiv (bT_0)^{2/3}/a^\prime c^{1/3}$
be sufficiently small that
\begin{equation}
\Delta t_{1D} \ll \left( { J \over a^\prime} \right)^{2/3}.
   \label{ad10}
\end{equation}
If this is not the case one can integrate out
the one-dimensional fluctuations to derive a new Ginzburg-Landau
functional with renormalized coefficients,
\begin{equation}
\tilde F[\Phi(x,y,z)]= {1 \over a_x a_y}\int d^3 x \left[
A \mid \Phi \mid^2
+B \mid \Phi \mid^4
+ C_x \mid {\partial \Phi \over \partial x }\mid^2
+C_y  \mid {\partial \Phi \over \partial y }\mid^2
+ C_z \mid {\partial \Phi \over \partial z }\mid^2
\right]
\label{bg1}
\end{equation}
where $a_x$ and $a_y$ are the lattice constants
perpendicular to the chains.
The new order parameter $\Phi(x,y,z)$, is proportional
to the average of
$\phi_i(z)$ over neighbouring chains.
The three-dimensional mean-field temperature $T_{3D}$ is defined
as the temperature at which the the coefficient $A(T)$
changes sign.
 Close to $T_{3D}$
\begin{equation}
A=A^\prime\left({T \over T_{3D}} -1 \right).
\label{abg1}
\end{equation}

The transition temperature $T_{3D}$ and the
coefficients $A^\prime$, $B$, $C_x$, $C_y$, and $C_z$
can be written in terms of the interchain interaction
$J$ and  the coefficients
$a$, $b$, and $c$ of a single chain.
The coefficients in (\ref{bg1}) determine measurable quantities associated
with the transition such as the specific heat jump,
coherence lengths and width of the critical region.

Most of the physics is
determined by a {\it single} dimensionless parameter
\begin{equation}
\kappa  \equiv { 2 (bT)^2 \over |a|^3 c}.
\label{aat1}
\end{equation}
which is a measure of the fluctuations along a single chain.
It was assumed that the coefficients $a$, $b$, and $c$
were independent of temperature and the measurable
quantities at the transition were determined as a function
of the interchain coupling. The transition temperature
increases as the interchain coupling increases. The coherence
length and specific heat jump depends only on the
single chain coherence length, $\xi_0 \equiv (c /|a|)^{1/2}$,
and the interchain coupling.
The width of the critical region, estimated from the
Ginzburg criterion, was virtually parameter independent,
being about 5-8 per cent of the transition temperature for
a tetragonal crystal. Such a narrow critical region is
consistent with experiment, and shows that Ginzburg-Landau
theory should be valid over a broad temperature range.

This paper uses  a simple model to demonstrate
some of the difficulties involved in deriving the coefficients
$a$, $b$, $c$, and $J$ from a realistic microscopic theory.

\subsection{Microscopic theory}

The basic physics of quasi-one-dimensional CDW
materials is believed to be described by a Hamiltonian due
to Fr\"ohlich \cite{fro} which describes electrons
with a linear coupling to phonons.
Even in one dimension this is a highly non-trivial
many-body system and must treated by some approximation scheme.
The simplest treatment \cite{fro,ric0,all} is a rigid-lattice one
in which the phonons associated with the lattice
distortion are treated in the mean-field approximation
and the zero-point and thermal lattice motions are neglected.
The resulting
theory is mathematically identical to BCS theory \cite{all}.
An energy gap opens
at the Fermi surface at a temperature
$T_{RL} \simeq 1.14E_F e^{-1/ \lambda}$ where $E_F$ is
the Fermi energy and $\lambda$ is the dimensionless
electron-phonon coupling.
$T_{RL}$ is  related to the
zero-temperature energy gap $\Delta_{RL}(0)$ by
\begin{equation}
 \Delta_{RL}(0) = 1.76 k_B T_{RL}.
\label{trl}
\end{equation}
In this approximation the coefficients in the single-chain
Ginzburg -- Landau free energy functional (\ref{aa1}) are \cite{all}
\begin{equation}
a_{RL}(T)= {1 \over \pi v_F} \ln \left({T \over T_{RL}}\right)
\label{mfa}
\end{equation}
\begin{equation}
b_{RL}(T)= {1 \over \pi v_F} {7\zeta(3)\over(4\pi T)^2}
\label{mfb}
\end{equation}
\begin{equation}
c_{RL}(T)={1 \over \pi v_F}{7\zeta(3)v_F^2\over(4\pi T)^2}
\label{mfc}
\end{equation}
where $v_F$ is the Fermi velocity and $\zeta(3)$ is
the Riemann zeta function.
If $4t_\perp$ is the electronic bandwidth perpendicular to the
chains (see (\ref{fk}))
 then the interchain coupling is given by \cite{sch,hor}
\begin{equation}
J_{RL}=  \left( {4 t_\perp \over v_F}\right)^2 c_{RL}(T).
\label{mfc2}
\end{equation}

It might be hoped that the transition in real materials
can be described by the mean-field  theory of the
functional (\ref{aa1}) with the coefficients (\ref{mfa}-\ref{mfc}).
However, this is not the case for several reasons.
(i) The width of the critical regime given by the one-dimensional
Ginzburg criterion \cite{ma} is very large:
 $\Delta t_{1D} = 0.8 $ \cite{sch}, suggesting that
fluctuations are important because condition (\ref{ad10}) is
not satisfied.
(ii) A rigid-lattice treatment predicts a metallic density of
states at all temperatures above $T_{RL}$.
In contrast, magnetic susceptibility \cite{sco,joh,joh3},
 optical conductivity \cite{deg,deg2,dre,dre2,bru,ber},
and photoemission \cite{dar,dar2,hwu} measurements suggest
 that there is a gap
or pseudogap in the density of states for a broad temperature range
above $T_P$.
(iii) The transition temperature, specific heat jump, and coherence
lengths are inconsistent with rigid lattice predictions
 (Table \ref{table1}).
This failure should not be surprising given that recent work has
shown that in the three-dimensionally ordered Peierls state
the zero-point and thermal lattice motions must
be taken into account to obtain a quantitative
description of the optical properties \cite{deg,deg2,mck,kim,lon}.

The next level of approximation is to use the coefficients
(\ref{mfa}-\ref{mfb}) and take into account the intrachain
order parameter fluctuations and the interchain coupling
and use results similar to those in References \cite{mck2}.
This is the approach that has been taken previously
\cite{sch,lee,die}.
There are two problems with this approach. First, if
the dimensionless parameter $\kappa$, given by (\ref{aat1}), is evaluated
using the expressions (\ref{mfa}-\ref{mfc})
the result is
\begin{equation}
\kappa_{RL}(T)=  { 7 \zeta(3) \over 8 |\ln (T/T_{RL})|^3 }.
\label{at1}
\end{equation}
Hence, the temperature dependence is quite different
from the dependence
$\kappa \sim T^2$ that was assumed in References \cite{mck2,die,sca2}
and the analysis there needs to be modified.
The second and more serious problem is
one of self-consistency. The coefficients $a$, $b$, and $c$
are calculated neglecting fluctuations in the order
parameter which will modify the electronic properties
which in turn will modify the coefficients.
In this paper a simple model is used to demonstrate that
the fluctuations have a significant effect on the
single-chain coefficients.

An alternative microscopic theory, due to Schulz \cite{sch},
and which takes into account fluctuations in only the phase
of the order parameter
is briefly reviewed in Appendix \ref{appsch}.

\subsection{Overview}

Discrepancies between phonon rigid-lattice
theory and the observed properties of the Peierls state well
below the transition temperature $T_P$
were recently resolved \cite{mck,kim}  by taking into account the
effect of the zero-point and thermal lattice motion on
the electronic properties. It was shown that the lattice fluctuations
have an effect similar to a Gaussian random potential. This mapping
breaks down near the transition temperature because of the phonon dispersion
due to the softening of the phonons near $2k_F$. In this paper
this dispersion is taken into account and the effect of the
large thermal lattice motion near the transition temperature
is studied.

The thermal lattice motion has the same effect on the electronic properties
as a static random potential with finite correlation length.
Close to the transition temperature, the problem reduces to
a simple model, corresponding to a single classical phonon,
which can be treated exactly (Section \ref{secham}).
This model was first studied by Sadovsk\~i\~i \cite{sad}.
It was recently used in the description of the destruction of
spin-density-wave states by high magnetic fields \cite{mck0}.
The one-electron Green's function is calculated in Section \ref{secgreen}.
There is a pseudogap in the density of states
(Fig \ref{figdos}).
The complexity of this simple model is indicated by two non-trivial
many-body effects: (i) Perturbation theory diverges
but is Borel summable. (ii) The traditional quasi-particle
picture breaks down
(Figure \ref{figspec}),
 reminiscent of behaviour seen in Luttinger liquids\cite{voi}.
To illustrate that calculations based on perturbation
theory can be unreliable it is shown that a predicted scaling
relation between the specific heat and the temperature
derivative of the magnetic susceptibility \cite{cha} does not hold
if the {\it exact}, rather than approximate, density of
states is used in the calculation.
Using this model the coefficients $a$, $b$, and $c$
are calculated in Appendix \ref{seccoeff}.
The coefficients deviate significantly from the
rigid-lattice values  if the pseudogap is
 comparable to or larger than the transition temperature.
 In Section \ref{secest} experimental data
 is used to estimate the pseudogap in
K$_{0.3}$MoO$_3$.

\subsection{Previous work on fluctuations and the pseudogap}

To put this paper in context some important earlier work
is briefly reviewed.

Lee, Rice and Anderson \cite{lee}
considered how fluctuations in the order parameter
produce a pseudogap in the density of states.
It is important to be aware of the assumptions
made in their calculation. Although their results describe
much of the physics on a qualitative level, for the
reasons described below, their results cannot be expected to give
a quantitative description of the density of states near
the CDW transition.
The starting point of Lee, Rice, and Anderson
was the one-dimensional
Ginzburg-Landau functional (\ref{aa1}) with a {\it real}
order parameter and with  the coefficients derived from
rigid-lattice theory (see equations (\ref{mfa}-\ref{mfc})).
Earlier, Scalapino, Sears, and Ferrell \cite{sca}
evaluated the  correlation length $\xi_\parallel(T)$
for one-dimensional Ginzburg-Landau theory
with an exact treatment of the fluctuations in the
order parameter;
$\xi_\parallel(T)$ only diverges as $T \to 0$.
The results of this calculation were used by Lee, Rice, and Anderson as
input in a random potential
with correlations given by
\begin{equation}
<\Delta(z)\Delta(z')>=\Delta_{RL}
(T)^{2} \exp(-\mid z-z^{'}\mid /\xi_\parallel(T)) \label{af}
\end{equation}
where $\Delta_{RL}(T)$ is the rigid-lattice (BCS) order parameter
and the average is over the thermal fluctuations of the
order parameter.
  The electronic
Green's function was calculated using equation (\ref{af}) and a
formula originally used for liquid
metals (essentially, second-order perturbation theory
for the random potential).
They found a gradual appearance of a gap as the
temperature decreased. For  $T_P < 0.25 T_{RL},$ an absolute gap of
magnitude $\Delta_{RL}(0)$ appears.
Lee, Rice and Anderson
suggested that a three-dimensonal transition
occurs for $T_P \simeq 0.25 T_{RL}$ based on the
temperature at which $\xi_\parallel(T)$ becomes extremely large.
There are several problems with trying to use these results
to give a quantitative description of the CDW transtion,
because of the following assumptions.
(i) {\it A  real order parameter.}
Most CDW transitions are described by a complex order parameter,
for which quantitatively distinct behaviour occurs.
For example, the transition to very large correlation
lengths for  $T_P \simeq 0.25 T_{RL}$ does not occur
for a complex order parameter. (See Figure 6 in Reference
\cite{sca}).
(ii) {\it Rigid-lattice coefficients.}
It is shown in this paper that the pseudogap due to the
thermal lattice motion causes the Ginzburg-Landau coefficients to deviate
significantly from their rigid-lattice values (Figure \ref{figcoeff}).
(iii) {\it Perturbation theory.}
It is demonstrated in this paper that this is unreliable.
In particular as $\xi_\parallel(T) \to \infty $  in (\ref{af})
only a pseudogap rather than  an absolute gap
develops in the density of states (Figure \ref{figdos}).

Rice and Str\"assler \cite {ric} calculated the contribution of the phonon
fluctuations to the electronic self energy in the Migdal approximation,
i.e., second-order perturbation theory.
Interchain interactions were included through an anisotropic phonon
dispersion.  They found
 a pseudogap in the density of states
above the transition temperature. At $T_P$ there is an absolute gap
whose magnitude is determined by the electron-phonon coupling
and the interchain interactions.
They equated the observed transition temperature
with the single-chain mean-field transition temperature $T_0$
which they found to be significantly
reduced below the rigid-lattice value $T_{RL}$
 and to vanish as the interchain coupling vanishes.

In the limit of weak interchain interactions the analytic
form of the density of states
is identical to that of Lee, Rice, and Anderson \cite{lee}.
However, it is not commonly appreciated that the
origin  of the pseudogap in the two calculations is quite different.
The magnitude of the Rice and Str\"assler pseudogap
is proportional to the thermal lattice motion (compare Section \ref{sectherm}),
while the pseudogap studied by Lee, Rice, and Anderson pseudogap is
by assumption equal to the rigid-lattice gap $\Delta_{RL}(T)$.
Calculations similar to that of Rice and Str\"assler have been
performed by Bjeli\~s and Bari\~si\~c \cite {bje}, Suzumura and Kurihara
\cite {suz}, Patton and Sham \cite {sha}, and Chandra \cite{cha}.
The main problem with these calculations are that they are
based on perturbation theory.

\section {MODEL HAMILTONIAN}
\label{secham}

The starting point for this paper  is the
following one-dimensional model. The states in
an electron gas with Fermi velocity $v_F$
are described by spinors $\Psi(z)$. The upper and
lower components describe left and right moving electrons,
respectively.
The phonons are described by the field
\begin{equation}
\Delta(z) = g \sum_q \sqrt {\hbar \over 2M \omega_{2k_F+q}}
(b_{2k_F+q} + b_{-2k_F-q}^\dagger ) e^{iqz}
\end{equation}
where $b_s$ destroys a phonon of momentum $s$ and frequency $\omega_s$
and $g$ is the linear electron-phonon coupling.
The dimensionless electron-phonon coupling $\lambda$ is defined by
\begin{equation}
\lambda = 2 g^2 a_z/\pi v_F \omega_Q \label {bd}
\end{equation}
where $a_z$ is the lattice constant along the chains.
The electronic part of the Hamiltonian is \cite{bra}
\begin{equation}
H_{el} = \int dz \Psi^\dagger (z) \bigg[ - iv_F \sigma_3
{\partial \over \partial z} + {1 \over 2}(\Delta(z) \sigma_+ + \Delta(z)^*
\sigma_-)\bigg] \Psi(z)
\label{hamel}
\end{equation}
where $\sigma_3$ and
$ \sigma_{\pm} \equiv \sigma_1 \pm i \sigma_2$ are Pauli matrices.


This paper focuses on the following model where $\Delta(z)$
is replaced with a random potential with zero mean
and finite length correlations
\begin{equation}
\langle \Delta(z)\rangle = 0 \ \ \ \; \ \ \ \ \
\langle \Delta(z)\Delta(z')^* \rangle = \psi^2
\exp(-|z-z'|/\xi_\parallel).
\label{cor2}
\end{equation}
 $\xi_\parallel$ is the CDW correlation length along the chains.
In most of this paper $\psi$ will be treated as a parameter.
It is central to this paper, being
a measure of the thermal lattice motion and
a measure     of the pseudogap in the density of states.
This paper focuses on behaviour near $T_P$ and so the
limit $\xi_\parallel \psi/v_F \to \infty$ is taken.
A rough argument is now given to justify using this
model to describe thermal lattice motion near the phase transition.

\subsection {Thermal lattice motion}
\label{sectherm}

In rigid-lattice theory $\Delta(z)$ is replaced by its expectation value
$\langle \Delta(z) \rangle =\Delta_o$.
To go beyond this
the effect of the  quantum and thermal
lattice fluctuations in the Peierls state was
recently modelled \cite{mck,kim,mck1} by treating $\Delta(z)$
as a static random potential with mean $\Delta_o$ and correlations
\begin{equation}
\langle \Delta(z)\Delta(z')^* \rangle =\Delta_o^2 + \gamma \delta(z-z')
\label{corprl}
\end{equation}
where
\begin{equation}
\gamma= {1 \over 2}\pi \lambda v_F  \omega_{2k_F}
\coth\left({\omega_{2k_F} \over 2T}\right).
\label{gamma}
\end{equation}
This model is expected to be reliable except near the transition
temperature where there is significant dispersion in the phonons.
This dispersion is now taken into account.

Near the transition temperature the
phonons can be treated {\it classically} since in most materials the
frequencies of the phonons with wavevector near $2k_F$ are much smaller than
the transition temperature (Table \ref{table2}).
Following Rice and St\"assler \cite{ric} renormalized phonon frequencies
$\Omega(q,T)$ are used  in
the expression for the correlations of the random potential
\begin{equation}
\langle \Delta(z)\Delta(z')^* \rangle =
\lambda\pi T{ v_F\over a_z}
\sum_q{\omega_Q^2\over\Omega(q,T)^2}
e^{iq(z-z')}.
\label{cor}
\end{equation}
At the level of the Gaussian approximation the
phonon dispersion relation can be written in the form
\begin{equation}
\Omega(q,T)^2=\Omega(T)^2 \bigl(1 + (q- 2k_F)^2
\xi_{\parallel}(T)^2 \bigr). \label{be1}
\end{equation}
Evaluating (\ref{cor}) then gives (\ref{cor2}) where
\begin{equation}
\psi^2=\lambda \pi T
\left({\omega_Q \over \Omega(T)}\right)^2 {v_F\over 2\xi_\parallel(T)}.
\label{ce}
\end{equation}
Note that this expression togehter with (\ref{cor2})
 is then quite different from (\ref{af})
used by Lee, Rice, and Anderson \cite{lee}.
In the limit $\xi_\parallel \to 0$, i.e., the phonons become
dispersionless and the sum in (\ref{cor}) becomes a delta function
and giving (\ref{corprl}) (with $\Delta_o =0$) and (\ref{gamma}).

The rms fluctuations $\delta u$ in the
positions of the atoms due to
thermal lattice motion is related to the Debye-Waller factor
and given by
\begin{equation}
(\delta u)^2=kT \sum_q {1\over M\Omega(\vec{q},T)^2} \label {ca}
\end{equation}
This is related to $\psi = (2M \omega_Q)^{1/2} g \delta u$.
Hence $\psi$ {\it is proportional to the thermal lattice motion.}

If $\psi$ is defined by (\ref{ce}) it
diverges as $T \to T_P$ because
\begin{equation}
{\rm as}\  T \to T_P,\ \Omega(T) \to 0,\ \xi_\parallel(T) \to
\infty \ {\rm with}\  \Omega(T)\xi_\parallel(T)  \
{\rm finite.} \label{bf}
\end{equation}
However, in a real crystal the phonons are three-dimensional
and the thermal lattice motion is finite.
Define
\begin{equation}
\psi^2=\lambda\pi T{ v_F\over a_z}
\sum_{\vec{q}}{\omega_Q^2\over\Omega(\vec{q},T)^2} \label{cd}
\end{equation}
and write the three-dimensional dispersion relation
(for a tetragonal crystal) in the form
\begin{equation}
\Omega(\vec q,T)^2=\Omega(T)^2 \bigl(1 + (q_{\parallel}- 2k_F)^2
\xi_{\parallel}(T)^2
      + (q_{\perp}- Q_{\perp})^2 \xi_{\perp}(T)^2 \bigr) \label{be}
\end{equation}
where $\vec Q =(Q_\perp,2k_F)$ is the nesting vector associated with
the three-dimensional CDW transition (see equation (\ref{bb})).
Due to the quasi-one-dimensionality of the crystal
the dispersion perpendicular to the chains is small and
$\xi_\perp \ll \xi_\parallel$.
Let $a_x$ denote the lattice constant perpendicular to the
chains.
Performing the integral over the wave vector in (\ref{cd}) gives
\cite{alternative}
\begin{equation}
 \psi^2 = \lambda \pi T
\left({\omega_Q\over\Omega(T)}\right)^2
{a_x^2 v_F \over \pi^2 \xi_\parallel(T)\xi_\perp(T)^2}
\left[\sqrt{1+(\rho_c\xi_\perp(T))^2}-1\right]
\label{cf}
\end{equation}
where $\rho_c$ is a wavevector cutoff perpendicular to the chains.
If  $\rho_c=\pi/a_x$ this expression reduces to (\ref{ce}) in the
one-dimensional limit $\xi_\perp \ll a_x$.
Near the transition, $\xi_\perp(T) \to \infty$, giving
\begin{equation}
\psi(T_P)^2 = \lambda \pi T
\left({\omega_Q\over\Omega(T)}\right)^2 {a_x v_F \over \pi
\xi_\parallel(T)\xi_\perp(T) }\label{cg}
\end{equation}
 From (\ref{bf}) and the fact that $\xi_\parallel(T)/\xi_\perp(T)$
is finite it follows that $\psi$ is finite as $T \to T_P$.
Note that the magnitude of this quantity is dependent on the
choice of the momentum cutoff $\rho_c$.
The above treatment is quite similar to Schulz's discussion of
fluctuations in the order parameter in the Gaussian approximation
\cite{sch}.

Although the expressions (\ref{ce}), (\ref{cf}), and
(\ref{cg}) for $\psi$ in the different regimes look very
different $\psi$ is actually weakly temperature dependent and does not
vary much in magnitude. To see this (\ref{cf}) can be written as
\begin{equation}
\psi^2 = (\psi(T_P))^2
{1 \over \rho_c\xi_\perp(T)}\left[\sqrt{1+(\rho_c\xi_\perp(T))^2}-1\right]
\label{cg2}
\end{equation}
The postfactor is a slowly varying function of $\rho_c\xi_\perp(T)$.
Since well above $T_P$, $\rho_c\xi_\perp(T) \sim 1$
(e.g., for K$_{0.3}$MoO$_3$ $\xi_\perp(300 {\rm K}) \sim 4 \AA $ \cite{gir})
the postfactor does not vary by more than a factor of two
although $\rho_c\xi_\perp(T)$ varies by several orders of
magnitude.
Johnston et al. \cite{joh} used a crude method of estimating the
pseudogap and found it to be weakly temperature dependent above $T_P$
for K$_{0.3}$MoO$_3$.

\subsection{Solution of the model}
\label{secsol}

Sadovsk\~i\~i \cite{sad2} solved the one-dimensional model (\ref{hamel})
and (\ref{cor2}) exactly. He calculated the one-electron Green's function
in terms of a continued fraction by
finding a recursion relation satisfied by the self energy.
He found \cite{sad}
 that the Green's function reduced to a simple analytic form
in the limit of large correlation lengths ($\xi_\parallel \gg v_F/\psi$).
This can be seen by the following rough argument.
In the limit $\xi_\parallel \to \infty$
the moments of the random potential $\Delta(z)$ are independent of
position:
\begin{equation}
\langle \Delta(z)\rangle = 0
\ \ \ \ \ \ \langle \Delta(z)\Delta(z')^* \rangle = \psi^2.
\label{cori}
\end{equation}
This means that
the random potential has only one non-zero Fourier component, i.e.,
the one with zero wavevector.

The potential can be written $\Delta(z)=v \psi$
where $v$ is a complex random variable with a Gaussian distribution.
Averages over the random potential can then be written
\begin{equation}
\langle A[\Delta(z)]\rangle=
\int{dvdv^*\over\pi} e^{-vv*} A[v \psi].
\end{equation}
It is then a straight-forward exercise
to evaluate the averages of different electronic
Green's functions.


\section {ONE-ELECTRON GREEN'S FUNCTION NEAR $T_P$}
\label{secgreen}

The matrix Matsubara Green's function, defined at the Matsubara energies
$\epsilon_n=(2n + 1) \pi T$, for the Hamiltonian (\ref{hamel})
 with (\ref{cori}) is
\begin{equation}
 \hat G\left(i\epsilon_n ,k\right)=\int{dvdv^*\over\pi} e^{-vv*}
\hat G \left(i\epsilon_n,k,v\right)\label{ea}
\end{equation}
where
\begin{equation}
\hat G\left(i\epsilon_n,k,v\right)={-(i\epsilon_n
-kv_F\sigma_3-\psi(v\sigma_++v^*\sigma_-))\over\epsilon_n^2
+(kv_F)^2+vv^*\psi^2}\label{ea2}
\end{equation}
is the matrix Green's function for the Hamiltonian (\ref{hamel})
with $\Delta(z)=v\psi$.
The off-diagonal
(anomalous) terms vanish when the integral over $v$ is performed
indicating there is no long range order.  The integral over
the phase of $v$ can be performed  and variables
changed to $\varphi=vv^*$ and
obtain
\begin{equation}
\hat G(i\epsilon_n,k)=-(i\epsilon_n
-kv_F\sigma_3)\int_0^\infty d\varphi{e^{-\varphi}\over\epsilon_n^2
+(kv_F)^2+\varphi\psi^2}\label{eb}
\end{equation}
Sadovsk\~i\~i \cite {sad} obtained the same expression by
diagrammatic summation. For the case of a half-filled band $v$ is strictly
real and the resulting expressions are the same as those obtained by
Wonneberger and Lautenschl\"ager \cite{won}.
Expanding (\ref{eb}) in powers of $\psi$ gives
\begin{equation}
\hat G(i\epsilon_n,k)=\hat G_0(i\epsilon_n,k)\int_0^\infty d\varphi
e^{-\varphi}
\sum^\infty_{n=0}\left[{-\varphi\psi^2\over\epsilon_n^2+(kv_F)^2}
\right]^n \label{ec}
\end{equation}
where $\hat G_0=(i\epsilon_n -kv_F\sigma_3)^{-1}$
is the free-electron Green's function.  Performing the
integral over $\varphi$ gives
\begin{equation}
\hat G(i\epsilon_n,k)=\hat G_0(i\epsilon_n,k)\sum^\infty_{n=0}n!
\left[{-\psi^2\over\epsilon_n^2+(kv_f)^2}\right]^n \label{ed}
\end{equation}
This is a {\it divergent} series and asymptotic expansion.  However, it is
Borel summable \cite{bor}.  This divergence suggests that
perturbation theory as used in References \cite {lee,cha,ric,bje,suz,sha}
may give unreliable  results.
This can be seen in Figure \ref{figdos} and Section \ref{seccha}.

\subsection {Density of States}

The electronic density of states is calculated directly from the
imaginary part of the one-electron Green's function (\ref{eb}).  The result is
\begin{equation}
\rho(E)=\rho_o \int_0^\infty d\varphi
e^{-\varphi}{E\over\left[E^2-\varphi\psi^2\right]^{1/2}}
\theta\left(\mid E\mid^2-\varphi\psi^2\right)
\end{equation}
\begin{equation}
=2\rho_o
\bigl|{E\over\psi}\bigl|\exp(-\left({E\over\psi}\right)^2)
{\rm erfi}\left({E\over\psi}\right)\label{rf}
\end{equation}
where $\rho_o=1/\pi v_F$ is the free-electron density of states
and erfi is the error function of imaginary argument
\cite{err}.  Figure \ref{figdos}
 shows the energy dependence of the density of states.  It vanishes at
zero energy (the Fermi energy) and is suppressed over an energy
range of order $\psi$, i.e., there is a pseudogap.
It has the asymptotic behavior:
\begin{equation}
\rho(E) \simeq 2 \rho_o ({E\over\psi})^2 \quad {\rm for}
\ E\ll\psi \label{eg}
\end{equation}
$$\rho(E) \simeq \rho_o \quad  {\rm for} \ E\gg\psi \label{eg2}$$
Figure \ref{figdos}
 shows that the exact result (\ref{rf}) (solid line)
 deviates significantly from the result of
second-order perturbation theory in References \cite{lee,ric} (dashed line),
\begin{equation}
\rho(E)=\rho_o {E\over\left[E^2-\psi^2\right]^{1/2}}
\theta\left(E^2-\psi^2\right) \label{eh}
\end{equation}
This latter form has been assumed in much earlier
work \cite{hor,joh,sha,joh2}.

The above expressions for the density of states
are all for infinite correlation length
($\xi_\parallel \psi/v_F \to \infty $),
i.e., very close to the three-dimensional transition temperature $T_P$.
What happens
{\it above} $T_P$ as the intrachain correlation length decreases?
This problem was considered in detail by Sadovsk\~i\~i \cite{sad2}.
(He calculated the density of states for
the random potential
(\ref{cor2}) with finite $\xi_\parallel$  exactly).
As the correlation length decreases the density of states
at the Fermi energy increases, i.e., the pseudogap fills in.
How quickly this happens depends on the dimensionless
ratio $v_F/(\psi \xi_\parallel).$
(See equation (\ref{suc}) below and Figures 5 and 6 in Reference \cite{sad2}).
Sadovsk\~i\~i showed that perturbation theory \cite {lee,cha,ric,bje,suz,sha}
only gives reliable results for $|E| < \psi$
when $\xi_\parallel < v_F/\psi$, i.e.,
well above $T_P$.

What happens
{\it below} $T_P$ as the intrachain correlation length decreases?
 In Reference \cite{mck}  it was shown that in the
three-dimensionally ordered Peierls state,
well below $T_P$,
there is an absolute gap with a subgap tail that
increases substantially
as the temperature becomes larger than the phonon frequency.
A smooth crossover to the pseudogap discussed here is expected.
It is an open problem to construct a single theory that
can describe the density of states over the complete
temperature range.

\subsection {Spectral Function}

The spectral function for right moving electrons
of momentum $k$  is given by
\begin{eqnarray}
 A(k,E)&=&-{1\over\pi}{\rm Im}\ G_{11}(k,E+i\eta) \nonumber \\
&=&\int_0^\infty d\varphi e^{-\varphi}\left[
\delta\left(E-\sqrt{(kv_F)^2+\varphi\psi^2}\right)
+\delta\left(E+\sqrt{(kv_F)^2+\varphi\psi^2}\right)\right]
\nonumber \\
&=&{\mid E\mid\over\psi^2}\exp\left(
{(kv_F)^2-E^2\over\psi^2}\right)
\theta\left(E^2-(kv_F)^2\right) \label{ej}
\end{eqnarray}
where the momentum $k$ is relative to the Fermi momentum $k_F$.
Note that this form is very different from the Lorentzian form
associated with the quasi-particle picture and perturbation theory
\cite{rick}.
The spectral function
is asymmetrical, very broad, and has a significant high energy tail.
Figure \ref{figspec} shows how the quasi-particle weight is reduced
near the Fermi momenta, i.e., the quasi-particles are not well defined.
This was first pointed out by
Wonneberger and Lautenschl\"ager \cite{won} for the
corresponding model for a half-filled band.
This is strictly a non-perturbative effect. In perturbation theory
the quasi-particles are well defined.
This breakdown of the quasi-particle picture is similar to the
properties of a Luttinger liquid \cite{voi}.

The momentum distribution function $n(k)$ at $T=0$ for right moving
electrons is given by
\begin{equation}
n(k)\equiv \int_{-\infty}^0 dE A(k,E)
= {1\over 2} \left[ 1 - \sqrt{\pi}
 \left({kv_F \over \psi} \right)
\exp \left( \left({kv_F \over \psi} \right)^2 \right)(1-
{\rm erf} \left({kv_F \over \psi} \right))
\right]
\label{ek}
\end{equation}
where  ${\rm erf}$ is the error function.
The inset to Figure \ref{figspec}
shows how the momentum distribution $n(k)$
at $T=0$ is smeared over a momentum range $\delta k \sim \psi/v_F$.
 The absence of a step at $k=k_F$  indicates that there
is no clearly defined Fermi surface.
However, this is {\it not} like in a Luttinger liquid,
but solely due to disorder. In fact, in an ordinary metal
with mean free path $\ell$ similar behaviour is seen;
disorder smears out $n(k)$ over a momentum range $\delta k \sim  1/\ell$.

\subsection{Electronic specific heat }

The electronic specific heat $C_e(T)$ is related to the density of
states $\rho(E)$ by
\begin{equation}
C_e(T) = - {4 \over T} \int_0^\infty dE E^2
\rho(E) {\partial f \over \partial E}
\label{spa}
\end{equation}
where $f(E)$ is the Fermi-Dirac distribution function.
In the absence of a pseudogap $C_e(T)={2 \pi^2 \over 3} \rho_0 T
 \equiv C_0(T).$
If the expression (\ref{rf}) is used for the density of
states in the presence of a pseudogap then $C_e(T)/C_0(T)$
only depends on $\psi/T$ and is shown in Figure \ref{figparam}.
A similar result was recently used \cite{mck0} to explain the temperature
dependence of the electronic specific heat near a spin-density-wave
phase boundary of the organic conductor (TMTSF)$_2$ClO$_4$.
Note that when $\psi \sim T$, $C_e(T)$ can be slightly larger
than $C_0(T)$ because $E^2 {\partial f \over \partial E}$
has a maximum near $ E \sim T$ and for $ E \sim \psi$,
$\rho(E)$ is larger than $\rho_0$ (Figure \ref{figdos}).

\subsection{Pauli Spin Susceptibility}
\label{secsusc}

The Pauli spin susceptibility $\chi(T)$ is related to the density of
states $\rho(E)$ by
\begin{equation}
\chi(T) = -\mu_B^2 \int_0^\infty dE \rho(E) {\partial f \over \partial E}
\label{sua}
\end{equation}
where $f(E)$ is the Fermi-Dirac distribution function
and $\mu_B$ is a Bohr magneton \cite{ash}.
In the absence of a pseudogap $\chi(T)=\mu_B^2 \rho_0 \equiv \chi_0$
which is independent of temperature.
If the expression (\ref{rf}) is used for the density of
states in the presence of a pseudogap then $\chi(T)/\chi_0$
only depends on $\psi/T$ and is shown in Figure \ref{figparam}.
This result will be used in Section \ref{secest} to provide an
estimate of the pseudogap in K$_{0.3}$MoO$_3$.

\subsection{Chandra's scaling relation}
\label{seccha}

The effect of thermal lattice fluctuations on the temperature
dependence of $\chi(T)$ was first considered
by Lee, Rice, and Anderson \cite{lee}. They argued that as the
temperature is lowered towards $T_P$ the intrachain
correlation length increases, more of a pseudogap opens in
the density of states and $\chi(T)$ decreases.
This problem was recently reconsidered  by Chandra \cite{cha}
who derived a scaling relation between the derivative
$ d \chi / d T$ and the specific heat $C_P$ in the critical region.
I now repeat the essential features of her argument.
She calculated the electronic self energy in the Born
approximation, taking into account the interchain interactions
and the finite mean free path of the electrons. She assumed
 that the pseudogap is much larger than the transition
temperature ($\psi \gg T_P$; it will be shown in Section \ref{secest}
that this is poor approximation for K$_{0.3}$MoO$_3$) so that
$\chi(T) \simeq \mu_b^2\rho(0)$. Chandra also assumed that the temperature
dependence of the density of states at the Fermi energy
is determined solely by the temperature dependence
of $\xi_\parallel(T)$. Moreover, based on the Born approximation,
she found
\begin{equation}
 \rho(0) \sim  {1 \over \xi_\parallel(T)}.
\label{sua2}
\end{equation}
Defining $t \equiv |T-T_P|/T_P$, then gives the scaling relation
\begin{equation}
{d \chi(T) \over dT } \sim {d \over dT } {1 \over \xi_\parallel(T)}
\sim {d \over dT } t^{1/2} \sim C_P
\label{sub}
\end{equation}
where use has been made of the temperature dependence of $\xi_\parallel(T)$
and $C_P$ in the Gaussian approximation \cite{ma}.

This same scaling relation was suggested earlier
by Horn, Herman and Salamon \cite{horn}. They claimed to
have found the critical exponent for $d \chi / d T$ to be --0.5 for TTF-TCNQ.
Kwok, Gr\"uner, and Brown \cite{kwo2}
claim to have observed a scaling between
$d (T \chi) / d T$ and $C_P$ within a 30 K
region about $T_P=183$ K  for K$_{0.3}$MoO$_3$.
However, Mozurkevich  has argued that the
Gaussian approximation is not valid in this
temperature range \cite{moz}.
Chung {\it et al.} \cite{chu} found that $d \chi / d T$
was comparable to $C_P$ when a background contribution
was subtracted from the latter.
Brill {\it et al.} \cite{bri} found that $\chi$
was proportional to the entropy (evaluated from
integrating the specific heat) between 140 and 220 K.
 (This is equivalent to
a scaling between $d \chi / d T$
and $C_P$). They show that this is
what is expected if $\chi$ and $C_P$ are derived from
a free energy functional in which the complete
magnetic field dependence is contained in the
field dependence of $T_P$.

Chandra's derivation of the  scaling relationship (\ref{sub}) is
 not valid. It depends on (\ref{sua2})
which is a direct result of the perturbative treatment
of the lattice fluctuations. The exact Green's function calculated
by Sadovsk\~i\~i \cite{sad2} gives different results. He found that for
$\xi_\parallel(T) \gg v_F/\psi$
\begin{equation}
{\rho(0) \over \rho_0} \simeq (0.54 \pm 0.01)
({v_F \over \psi \xi_\parallel(T)})^{1/2}
\label{suc}
\end{equation}
(see Figure 6 in Reference \cite{sad2})
rather than (\ref{sua2}).
This will give
\begin{equation}
{d \chi(T) \over dT } \sim t^{-3/4}
\label{sud}
\end{equation}
and so the scaling relation (\ref{sub}) does not hold.
It should be stressed that this result assumes $\psi \gg T_P$,
a condition that is poorly satisfied in most materials
(Section \ref{secest}).

\section{PROPERTIES OF THE GINZBURG-LANDAU COEFFICIENTS}
\label{secprop}

In Appendix \ref{seccoeff} the coefficients $a$, $b$, and $c$ in the
Ginzburg-Landau free energy (\ref{aa1}) describing the
Peierls transition are evaluated in the presence of the
random potential (\ref{cori}) which is used here to model the
thermal lattice motion. The calculation is based on a linked
cluster expansion similar to that used to derive
the Ginzburg-Landau functional for superconductors \cite{has}.
 The results are:
\begin{equation}
a(T)= {1 \over \pi v_F} \left[
\ln\left({T\over T_{RL}}\right)+\pi
T\sum_{\epsilon_n}\left({1\over\left|\epsilon_n\right|}-\int_0^\infty
d\varphi
e^{-\varphi}{\epsilon_n^2\over\left(\epsilon_n^2+
\varphi \psi^2\right)^{3/2}}\right) \right]
\label{gla}
\end{equation}
\begin{equation}
b(T)={ T \over 4 v_F}\sum_{\epsilon_n}
\int_0^\infty d\varphi e^{-\varphi}
\left( {\epsilon_n^2 \over
\left(\epsilon_n^2+ \varphi\psi^2 \right)^{5/2}}
-{5 \varphi(\psi \ \epsilon_n)^2 \over
\left(\epsilon_n^2+ \varphi \psi^2 \right)^{7/2}}
\right)
\label{glb}
\end{equation}
\begin{equation}
c(T)={v_F T\over 4}
\sum_{\epsilon_n}\epsilon_n^2\int_0^\infty{d\varphi
e^{-\varphi} \over (\epsilon_n^2+\varphi\psi^2)^{5/2}}\label{ff}
\label{glc}
\end{equation}
The integrals over $\varphi$ in the above
expressions can be written in terms of error functions and incomplete
gamma functions \cite{err}.  However, for both numerical and
analytical calculations it is actually more convenient to
use the expressions above.
As $\psi \to 0$ the above expressions
reduce to the rigid-lattice values (\ref{mfa}--\ref{mfc}).

{\it Single-chain mean-field transition temperature.}
 $T_0$ is determined by the temperature
at which the second-order Ginzburg-Landau coefficient (\ref{gla}) vanishes:
\begin{equation}
a(T_0)=0. \label{fk2}
\end{equation}
This defines  relations between $T_0/T_{RL}$ and
$\psi/T_0$, shown in Figure \ref{figcoeff}
(The inset shows $T_0/T_{RL}$ versus $\psi/T_{RL}$).
The pseudogap
suppresses the transition temperature.  At a crude level, this is
because in the presence of a pseudogap
 opening a gap due to a Peierls distortion causes a smaller
decrease in the electronic energy than in the absence of a pseudogap.
In most materials $T_P < 0.4 T_{RL}$ (Table \ref{table2})
and so the inset of Figure \ref{figcoeff} implies $\psi \sim T_{RL}$
which is comparable to the zero-temperature gap.
Rice and Str\"assler \cite{ric} found from second-order perturbation theory
that for $T_0 \ll T_{RL},$ $\psi \simeq 1.05 T_{RL}.$
Thus, the single-chain mean-field transition temperature
can be quite different from
$T_{RL}$, defined by (\ref{trl}), and often referred to
as the mean-field transition temperature,
and so no experimental signatures are expected at $T=T_{RL}$.

{\it Fourth-order coefficient.}
The ratio of the fourth-order coefficient $b$ to its rigid-lattice value
as a function of the ratio of the pseudogap $\psi$ to the
temperature is shown in Figure \ref{figcoeff}.
Note that $b$ is negative for $\psi/T > 2.7$.
This will change the nature of the phase transition.
One must then include the sixth-order term in the free energy.
If it is positive (I have calculated it and found it to be positive
for this parameter range)
then the transition will be {\it first order.}
A complete discussion of such a situation is given by
Toledano and Toledano \cite{tol}.
Imry and Scalapino have discussed the effect of
one-dimensional fluctuations for this situation \cite{imr}.
At the mean-field level there is a co-existence of phases
for the temperature range defined by
\begin{equation}
0 < a(T) < { b(T)^2 \over 3 d(T)}
\label{fm0}
\end{equation}
where $d(T)$ is the sixth-order coefficient.
Hysteresis will be observed
in this temperature range.
I recently suggested that the first-order nature of the destruction
by high magnetic fields of spin-density-wave states
in organic conductors is due to similar effects \cite{mck0}.
If at low temperatures the electron phonon coupling
$\lambda$ is varied then $\psi/T_{RL} \sim \lambda e^{1/\lambda}$.
According to the inset of Figure \ref{figcoeff} there will
be a critical coupling below which the CDW phase will
be destroyed. This transition will be first order.
It is interesting that Altshuler, Ioffe, and Millis \cite{alt}
recently obtained a similar result for a two-dimensional
Fermi liquid (with a quasi-one-dimensional
Fermi surface) using a very different approach.

However, it should be pointed out that when $b$ is small
corrections due to other effects such as a finite correlation
length and interchain coupling
will be important and could make $b$ positive.
It is unclear whether this unexpected behaviour is only a result
of the simplicity of the model or actually is relevant to
real materials. The
three-dimensional transition occurs when the parameter $\kappa$,
defined by (\ref{aat1}), becomes sufficiently small \cite{mck2}.
Generally this is assumed to be due to the temperature becoming sufficiently
low. However, I speculate that the transition
could alternatively be driven by $b$ becoming
sufficiently small.
The fact that $\psi \sim (2-3) T_P$ in K$_{0.3}$MoO$_3$
(Section \ref{secest}) is consistent with $b$ being small.

{\it The coefficient of the longitudinal gradient term}
 is given by (\ref{glc}).
It can be shown that $c(T)/c^{RL}(T)$ is a universal
function of $\psi/T$ (see Figure \ref{figcoeff})
and that the pseudogap reduces the
value of $c$.

{\it Interchain coupling.}
Consider a crystal with tetragonal unit cell of dimensions
$a_x \times a_x \times a_z$,
where the z-axis is parallel to the chains.  For a tight-binding model
the electronic band structure is given by the dispersion relation
\begin{equation}
E(k)=-2t_\perp(\cos(k_xa_x)+\cos(k_ya_x))-2t_\parallel \cos(k_z a_z).
\label{ba}
\end{equation}
Assume the band-structure is highly anisotropic, i.e., $t_\parallel\gg
t_\perp$.
The Fermi velocity $v_F$ is defined by $v_F=2t_\parallel  a_z \sin(k_F a_z)$.
 Horovitz, Gutfreund, and Weger \cite {hor} have shown that
imperfect nesting of the Fermi surface (i.e., $E(k) \simeq - E(k+Q))$
occurs for the nesting vector
\begin{equation}
\vec{Q}=(\pi/a_x,\pi/a_x,2k_F). \label {bb}
\end{equation}

To calculate the interchain coupling $J$ in the
 Ginzburg-Landau functional (\ref{ad1}) it   is assumed that
 the one-dimensional Green's function (\ref{ea2})
can simply be replaced
with the corresponding one with the anisotropic band structure, given by
equation (\ref{ba}).
The calculation
 is then essentially identical to the rigid-lattice calculation of Horovitz,
Gutfreund, and Weger \cite{hor} and so  only the result is given
(compare (\ref{mfc2})):
\begin{equation}
J   =  \left( { 4 t_\perp  \over v_F}\right)^2 c(T).
 \label{fk}
\end{equation}
Since the pseudogap reduces the value of the longitudinal
coefficient $c$ it will also reduce the interchain coupling.

\section{MEAN-FIELD THEORY OF A SINGLE CHAIN}

The single chain Ginzburg-Landau functional
with the coefficients discussed in the previous section
is now considered.
In particular it is shown that the one-dimensional fluctuations
can be much smaller than for the functional with the
rigid-lattice coefficients.
The first step is to consider the temperature dependence of
the second-order coefficient $a(T)$ near $T_0$,
the mean-field transition temperature.
This is difficult because to be realistic the
temperature dependence of the parameter $\psi$
must be included. This is done at a crude level
using the simple model based on the discussion of
thermal lattice motion in Section \ref{sectherm}.
This is then used to evaluate $a^\prime$, defined
by (\ref{aa10}), and needed to evaluate physical
quantities associated with the transition:
the specific heat jump, the coherence length,
and width of the critical region.

The jump in the specific heat at $T_0$ is
\begin{equation}
\Delta C_{1D}=
{(a^\prime)^2 \over 2 \ b  \ T_0}.
\label{ac1}
\end{equation}
An important length scale is the coherence length $\xi_0$,
defined by
\begin{equation}
\xi_0=\left({c \over a^\prime}\right)^{1/2}
\label{acd1}
\end{equation}
The one-dimensional
Ginzburg criterion \cite{gin} provides an estimate of the
temperature range, $\Delta T_{1D}$, over which critical fluctuations
are important.
\begin{equation}
\Delta t_{1D} \equiv {\Delta T_{1D} \over T_0}
= \left({b  \ T_0 \over a^{\prime 3/2} c^{1/2}}
\right)^{2/3}
= {1 \over (2 \xi_0 \Delta C_{1D})^{2/3}}
\label{acc1}
\end{equation}

\subsection {Self-consistent determination of the pseudogap}

At the level of the Gaussian approximation the
phonon dispersion is related to the Ginzburg-Landau
coefficients by
\begin{equation}
\Omega(q,T)^2=\lambda \omega_Q^2 \bigl(a(T) +
c(T) (q- 2k_F)^2 + Ja_x^2 (q_\perp - Q_\perp)^2 \bigr). \label{bz1}
\end{equation}
Hence the phonon dispersion depends on the pseudogap
$\psi$.
However, it was shown  in Section \ref{sectherm} that $\psi$ depends on the
dispersion. Hence, $\psi$ must be determined self-consistently.
Equation (\ref{cg}) gives the dependence of the pseudogap at $T_0$ on the
phonon dispersion.  Equation (\ref{ff}) gives the dependence of the
coefficient $c(T)$ on the pseudogap.
 These can be combined with (\ref{fk}) to give
\begin{equation}
 1= t_\perp \psi^2 \
 \sum_{\epsilon_n}
\epsilon_n^2\int_0^\infty {d\varphi e^{-\varphi}\over
(\epsilon_n^2+\varphi\psi^2)^{5/2}}.\label{fm}
\end{equation}
It follows that $\psi/T$ is a universal function of
$t_\perp /T$.

{\it Dependence of $T_0$ on the interchain interactions.}
The self-consistent equation for the pseudogap (\ref{fm})
can be solved simultaneously with the equations for $T_0$,
and (\ref{fk})
to give the single-chain mean-field
transition temperature as function of the interchain interactions.
The transition temperature is then a monotonic
increasing function of the interchain hopping.
A similar   procedure was followed by Rice and Str\"assler
\cite{ric}.
The transition temperature tends to zero as the interchain
coupling tends to zero, consistent with the fact that there
are no finite temperature phase transitions in a strictly one-dimensional
system \cite{lan}.

\subsection {Evaluation of $a'$}

It is now assumed that the temperature dependence of
the pseudogap $\psi$ is
given implicitly by equation
(\ref{fm}).
Implicit differentiation then gives
\begin{equation}
{d \over dT} \left( {\psi \over T} \right)=
{\psi \over 2 T^2}
{X(T) \over Y(T)}
\label{yz}
\end{equation}
where
\begin{equation}
X(T)=\sum_{\epsilon_n}\epsilon_n^2\int_0^\infty{d\varphi
e^{-\varphi} \over (\epsilon_n^2+\varphi\psi^2)^{5/2}}
\end{equation}
\begin{equation}
Y(T)=\sum_{\epsilon_n}\epsilon_n^2\int_0^\infty{d\varphi
e^{-\varphi} \varphi \over (\epsilon_n^2+\varphi\psi^2)^{5/2}}
\end{equation}
Note that since the right-hand side of (\ref{yz})
is positive
 $\psi/T$ is always an increasing function of temperature.
A lengthy calculation gives
\begin{equation}
a'={1 \over \pi v_F} \left( 1 + {3 \over 2} \psi^2 \pi T
\sum_{\epsilon_n}\epsilon_n^2\int_0^\infty{d\varphi
e^{-\varphi} \over (\epsilon_n^2+\varphi\psi^2)^{5/2}}
\right)
\end{equation}
This is large than the rigid-lattice value
$a'_{RL} \equiv 1/\pi v_F$. This enhancement
will enhance the specific heat jump (\ref{ac1}) and reduce the
coherence   length (\ref{acd1}).

\subsection {Specific heat jump}

The specific heat jump $\Delta C$ at the transition  temperature
is calculated from equation (\ref{ac1}). It is shown in Figure
\ref{figjump}.
Note that the jump is much larger than the
rigid-lattice value of $1.43\gamma T_P$.
The trend shown in  Figure \ref{figjump} can be explained by a
rough argument correlating the sizes of $\Delta C/ \gamma T_P$
and $\Delta(0)/k_B T_P$. Simply put, if $\Delta(0)/k_B T_P$ is large
then $\Delta(T)^2$ will have a large slope at $T_P$.
It has previously been noted
experimentally \cite{sat} that the order parameter has a
BCS temperature dependence with $\Delta(0)$
and $T_P$ treated as independent parameters.
Some theoretical justification was recently provided
for such a temperature dependence well away from $T_P$ \cite{mck}.
Close to $T_P$ the BCS form gives
\begin{equation}
\Delta(T) \simeq 1.74 \Delta(0)\left(1-{T \over T_P}\right)^{1/2}.
\end{equation}
Within a BCS type of framework
the specific heat discontinuity is given by \cite{tin}
\begin{equation}
\Delta C \sim -\rho_o {d \Delta^2 \over dT} \Big|_{T_P}
 =  3.03 \rho_o  {\Delta(0) ^2 \over T_P}.
\end{equation}
Using $\Delta(0)=1.76k_B T_{RL}$ and $\gamma= 2 \pi^2 \rho_o/3$
gives
\begin{equation}
{\Delta C \over 1.43 \gamma T_P} \sim \left({ T_{RL} \over T_P}\right)^2
\end{equation}
This simple argument gives the correct trend that as the
fluctuations increase the enhancement of the specific heat jump
increases.

\subsection {Width of the one-dimensional critical region}

The width of the one-dimensional critical region
is calculated from equation (\ref{acc1})
with the Ginzburg-Landau coefficients in the
presence of the pseudogap. It is shown in Figure
\ref{figjump}, normalized to the rigid-lattice value
$\Delta t_{1D}=0.8$.
The large reduction is very important because it
means that even for weak interchain coupling,
it may be possible for condition (\ref{ad10}) to
be satisfied and for a mean-field treatment of
a single chain functional, such as that used in
this section, to be justified.

\section{Estimate of the pseudogap in  K$_{0.3}{\rm MoO}_3$}
\label{secest}

Optical conductivity, magnetic susceptibility and
photoemission experiments all suggest that near $T_P = 183$ K
 there is a pseudogap in the density of states.

{\it Optical conductivity}. Sadovsk\~i\~i has calculated the
optical conductivity $\sigma(\omega)$ for the model introduced
 in Section \ref{secham}
\cite{sad}. For small frequencies  $\sigma(\omega)$
is linear in $\omega$ and has a peak at about $\omega \simeq 3 \psi$.
The data in References \cite{deg,dre} then implies
$\psi \sim $ 40 meV
and $\psi/T_P \sim 2.5$.
On a less rigorous level $\psi$ can be estimated based on the
analysis contained in the inset of Figure \ref{figcoeff}. If
 the single-chain mean-field transition temperature $T_0 < 0.4 T_{RL}$
then $\psi \sim T_{RL}$. Using the BCS relation (\ref{trl})
and the estimate $\Delta (0) \simeq 80 $ meV for the zero-temperature
gap from the optical conductivity \cite{deg} gives
$\psi \sim $ 45 meV and $\psi/T_P \sim 3$.

{\it Magnetic susceptibility.}
The data of References \cite{bri,joh} gives $\chi(T_P)/\chi(300 K)
\simeq 0.5 $.
Assuming that $\chi(300 {\rm K}) \simeq \chi_0$
and using Figure \ref{figparam} gives $\psi /T_P \sim 2.4$.

Note that all of the above three estimates for $\psi/T_P$
are consistent with one another and
are all in the regime where the fourth-order coefficent $b$
is small (Figure \ref{figcoeff}).

{\it Photoemission.}
Recent high resolution photoemssion measurements
 \cite{dar,dar2,hwu}
 on K$_{0.3}$MoO$_3$ and (TaSe$_4$)$_2$I
 have several puzzling features:
(1) There is a suppression of spectral weight over a large energy range
(of the order of 200 meV for K$_{0.3}$MoO$_3$)
near the Fermi energy.
(2) The spectrum is very weakly temperature dependent. The suppression
occurs even for $T \sim 2 T_P$.
(3) At $T_P$ the spectrum does not just shift near $E_F$, due to the
opening of the Peierls gap, but also at energies of order 0.5 eV
from $E_F.$

These features {\it cannot} be explained using the model
presented in this paper.
The photoemission data suggests that the pseudogap is
about $\psi \sim $ 130 meV. Clearly this estimate is inconsistent
with the estimates ($\psi \sim $ 40-50 meV)
 given above from the optical conductivity
and magnetic susceptibility.
Furthermore, in the model presented here
the pseudogap occurs only when $\xi_\parallel(T) \gg v_F/\psi$,
i.e., fairly close to $T_P$.
The temperature dependence of the Pauli spin susceptibility
and the optical conductivity \cite{deg,deg2} suggest that
the pseudogap disappears for $T ~> 2 T_P$ (in contrast to
(2) above). Dardel et al. \cite{dar} speculate that the
anomalous   behaviour that they observe may arise
because the photoemission intensity $I(E)$ might be related
to the density of states $\rho(E)$ by $I(E)=Z \rho(E)$
and the quasi-particle weight $Z$ vanishes
due to Luttinger liquid effects.
This suggestion has been examined critically by Voit \cite{voi}
who concludes
that the photoemission data is only quantitatively
 consistent with a Luttinger
liquid picture if very strong long-range interactions are involved.
Kopietz, Meden, and  Sch\"onhammer \cite{kop}
have recently considered such models.

\section{CONCLUSIONS}

In this paper a simple model has been used to illustrate
some of the difficulties involved in  constructing from
microscopic theory a Ginzburg-Landau theory of the CDW transition.
The main results are:
(1) The large thermal lattice motion near the transition temperature
produces a pseudogap in the density of states.
(2) Perturbation theory diverges and gives unreliable results.
This is illustrated by showing that a predicted \cite{cha} scaling relation
between the specific heat and the temperature derivative
of the susceptibility does not hold.
(3) The pseudogap significantly alters the coefficients in the
Ginzburg-Landau free energy.
The result is that one-dimensional order parameter
fluctuations are less important,
making a mean-field treatment of the
single-chain Ginzburg-Landau functional more reasonable.

This work raises a number of questions and opportunities for
future work.
(a) The most important problem is that there is still no
microscopic theory that can make reliable quantitative
predictions about how dimensionless ratios such as
$\Delta (0)/k_B T_P$, $\Delta C/ \gamma T_P$, and $\xi_{0z}T_P/v_F$
depend on parameters such as $v_F$, the electron phonon coupling
$\lambda$, $T_P$ and the interchain coupling.
(b) Is the change of the sign of the fourth-order
coefficient $b$ of the single-chain Ginzburg-Landau functional
 for $\psi > 2.7 T_P$
an important physical effect or merely a result of
the simplicity of the model considered here?
(c) Calculation of the contribution of the sliding CDW
to the optical conductivity in the presence of the short-range
order associated with the pseudogap \cite{dre,dre2}.

\acknowledgments

I have benefitted from numerous discussions with J. W. Wilkins.
This work was stimulated by discussions with J. W. Brill.
I am grateful to him for showing me his group's data
prior to publication.
I thank K. Bedell and K. Kim for helpful discussions.
Some of this work was performed at The Ohio State University
and  supported by the U.S. Department of Energy,
Basic Energy Sciences, Division of Materials Science
and the OSU Center for Materials Research.
Work at UNSW was supported by the Australian Research Council.

\twocolumn
\narrowtext

\narrowtext
\twocolumn
\centerline{\epsfxsize=7.0cm \epsfbox{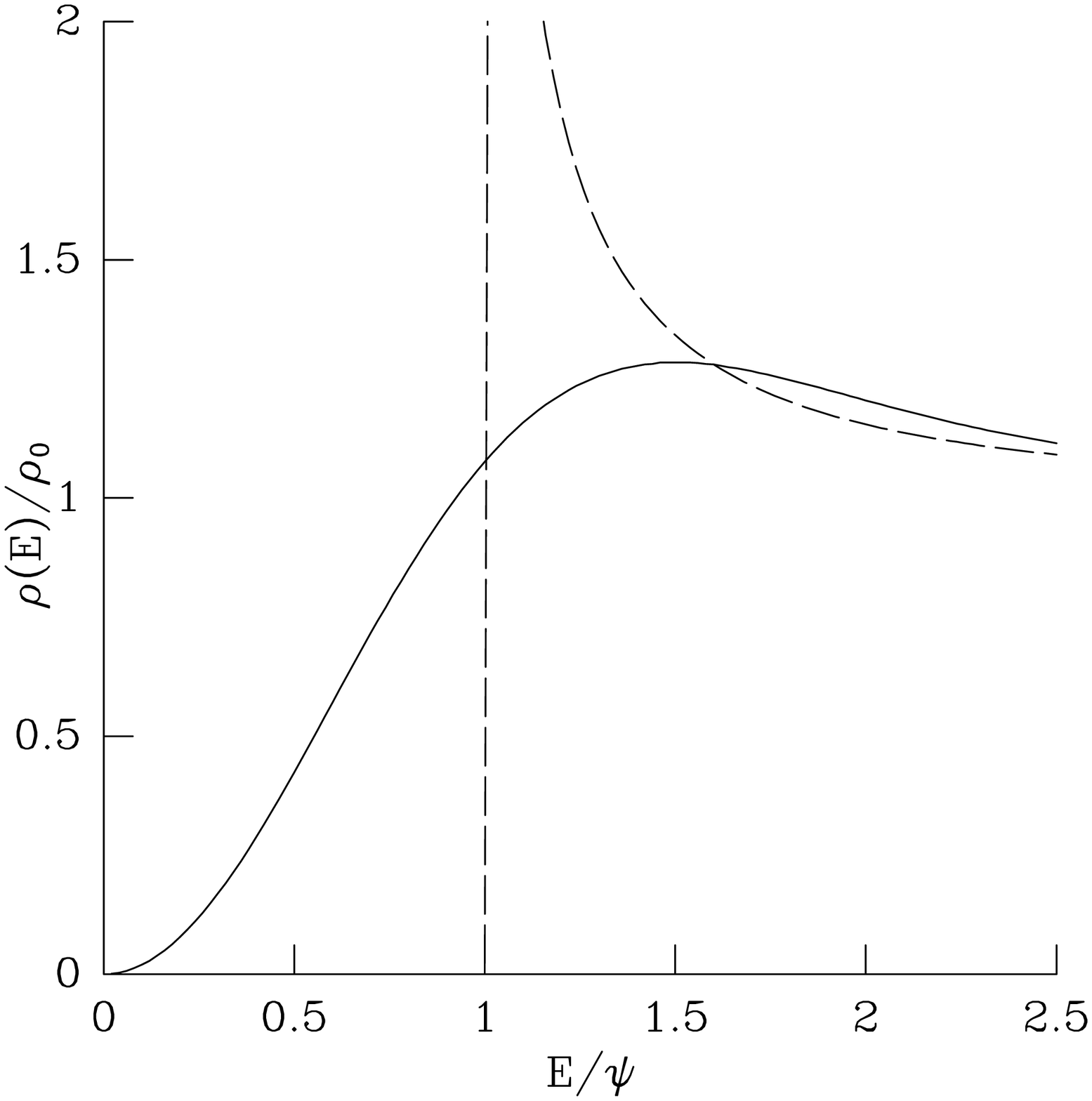}}
\begin{figure}
\caption{Pseudo-gap in the density of states near the
three-dimensional transition
temperature $T_P$. Perturbative treatments (dotted line,
compare Ref. \protect\cite{lee,ric}) give an absolute gap
$\psi$ at the transition temperature
whereas the exact treatment (solid line) gives
only a pseudogap. The energy $E$ is relative to the
Fermi energy and the density of states is normalized
to the free-electron value $\rho_o$.
The density of states is symmetrical about $E=0$.
This result is only valid sufficiently close to
$T_P$ that the longitudinal CDW correlation length
$\xi_\parallel \gg v_F/\psi$. As the temperature
increases above $T_P$, $\xi_\parallel $ decreases and
the density of states at the Fermi energy increases, i.e.,
the pseudogap gradually fills in (see Figures 5 and 6
in Reference \protect\cite{sad2}).
\label{figdos}}
\end{figure}

\centerline{\epsfxsize=7.0cm \epsfbox{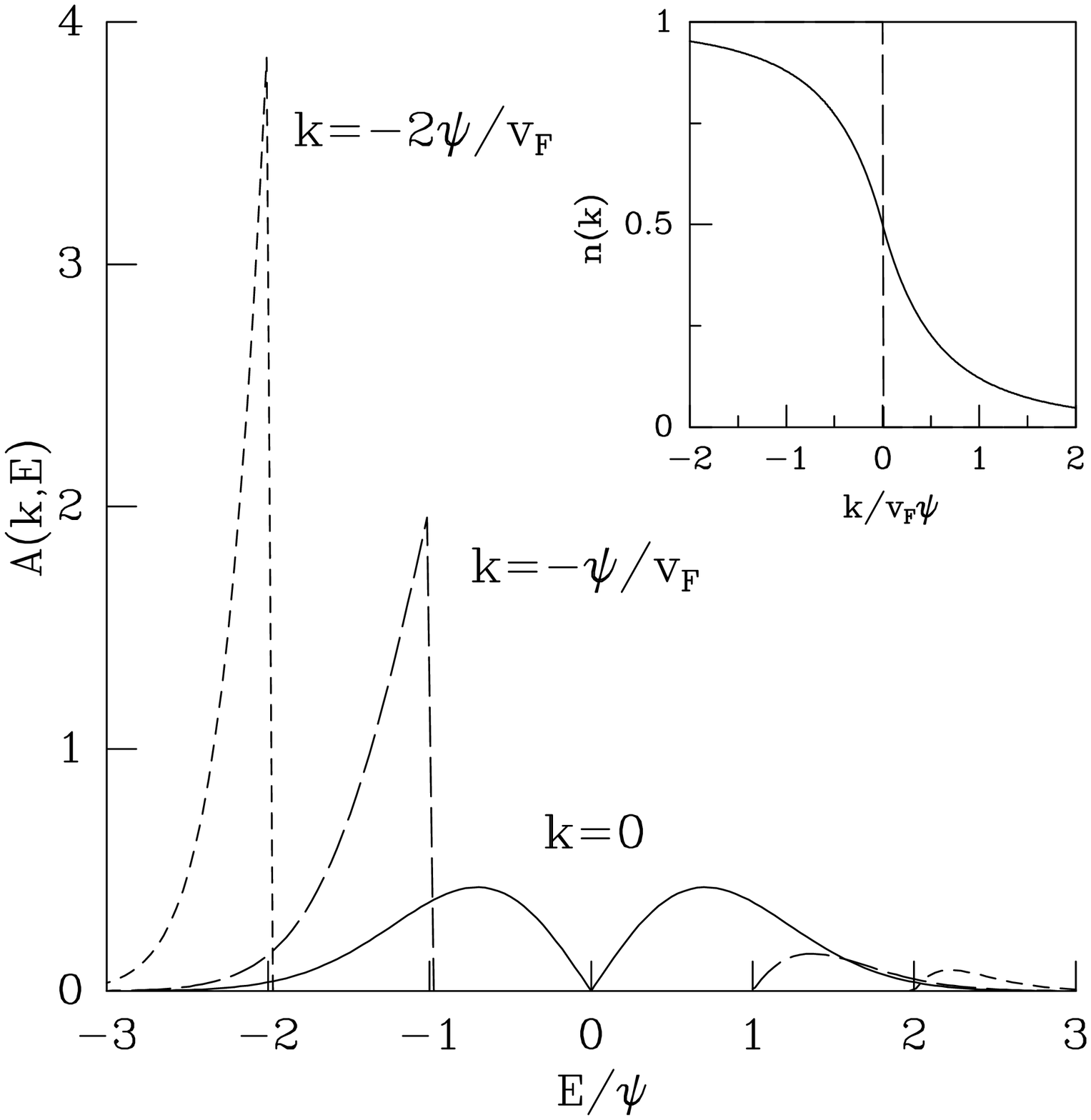}}
\begin{figure}
\caption{Breakdown of the quasi-particle picture. The electronic
spectral function is shown for several different momenta $k$,
relative to the Fermi momentum $k_F$.
As the momentum approaches $k_F$
the spectral function broadens
significantly, similar to the behaviour of a Luttinger liquid.
Inset: Momentum dependence of the occupation function $n(k)$.
The dashed line is the result in the absence of a pseudogap,
i.e., a non-interacting Fermi gas.
\label{figspec}}
\end{figure}

\centerline{\epsfxsize=7.5cm \epsfbox{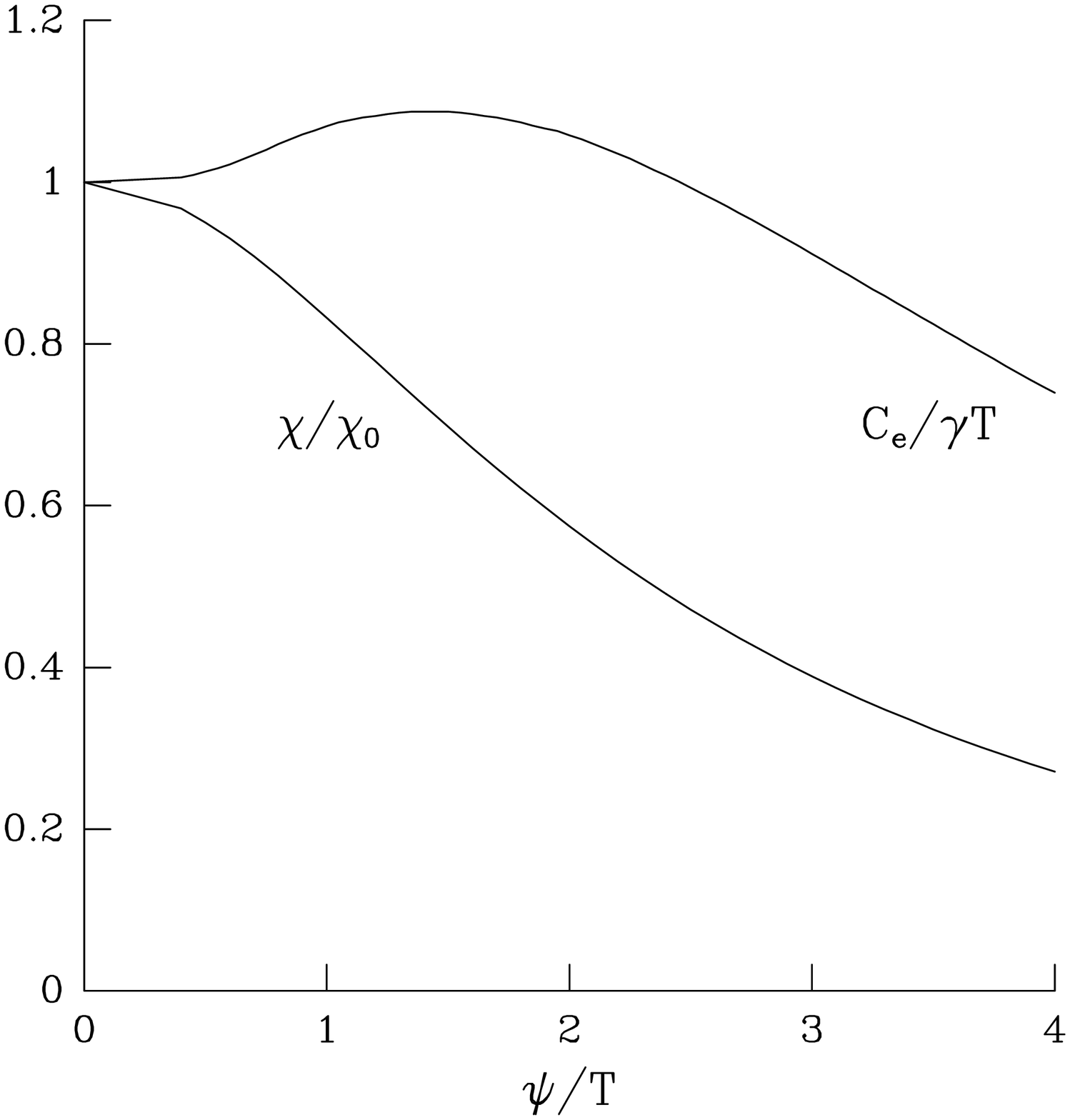}}
\begin{figure}
\caption{
Modification of the electronic specific heat $C_e(T)$ and
the Pauli spin susceptibility $\chi(T)$
by the pseudogap.
 Both are normalized to their values in the
absence of the pseudogap.
\label{figparam}}
\end{figure}

\centerline{\epsfxsize=7.5cm \epsfbox{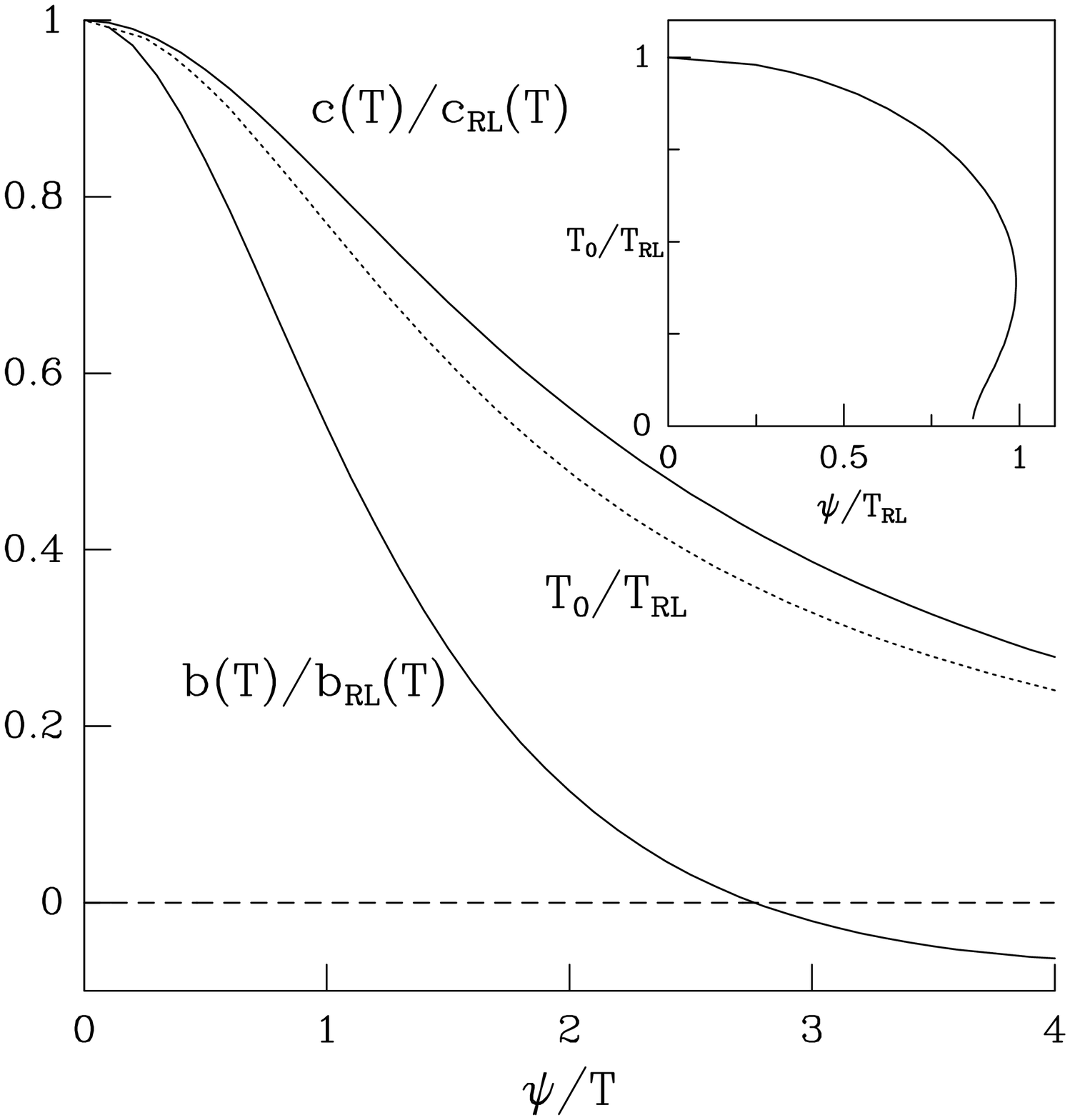}}
\begin{figure}
\caption{The pseudogap due to thermal lattice motion has
a significant effect on the coefficients in the
Ginzburg-Landau free energy (\protect\ref{aa1}) for a single chain.
The ratio of the single-chain
mean-field transition temperature $T_0$
and the coefficients $b$ and $c$ to
their rigid lattice values (given by (\protect\ref{mfa} -
\protect\ref{mfc})) are shown as a function of
the ratio of the pseudogap $\psi$ to the temperature.
For $\psi > 2.7 T$ the coefficient $b$ becomes
negative and the transition will be first order
(Section \protect\ref{secprop}).
Inset: Relationship between $T_0/T_{RL}$ and $\psi/T_{RL}$.
\label{figcoeff}}
\end{figure}

\centerline{\epsfxsize=7.5cm \epsfbox{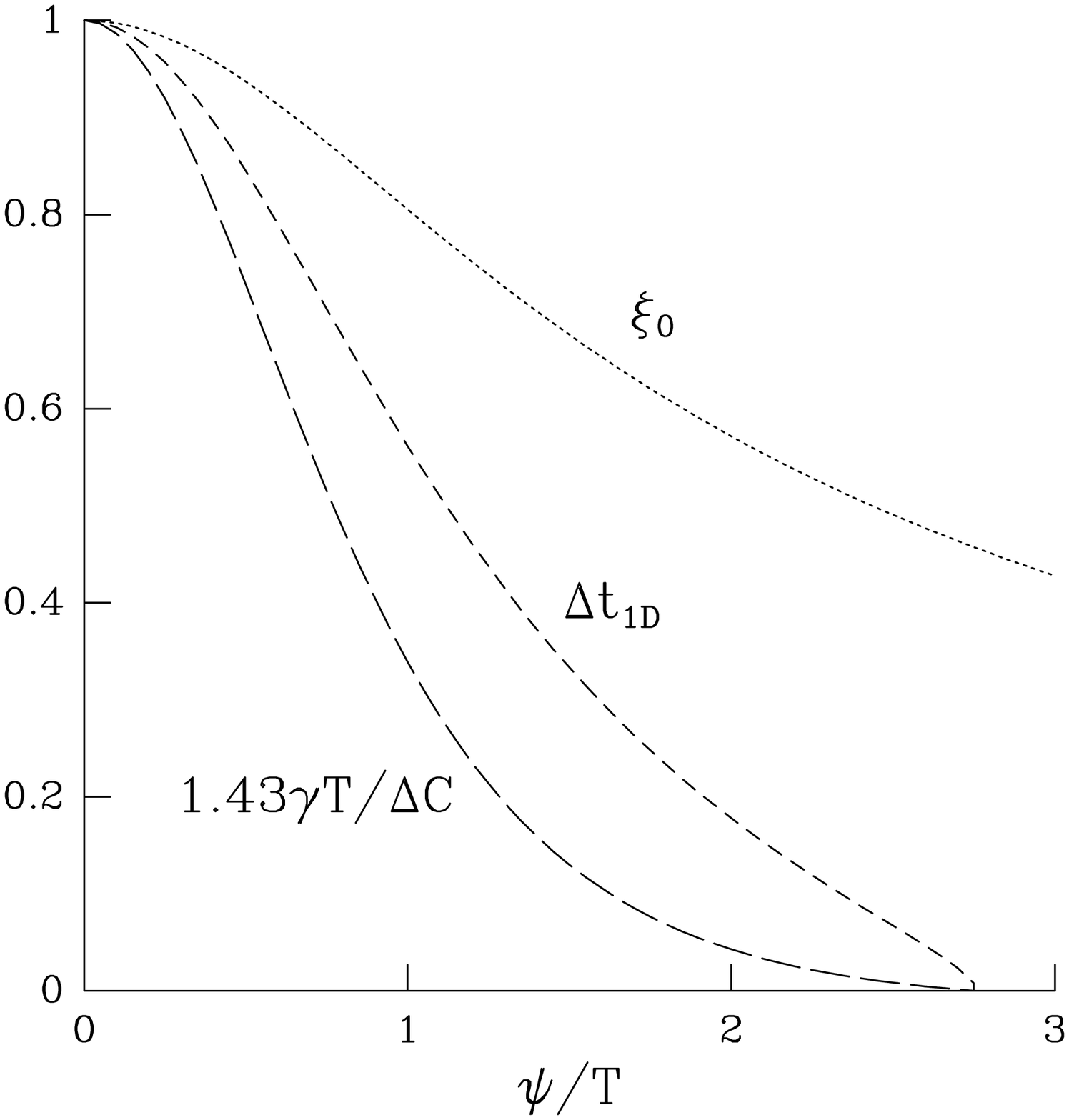}}
\begin{figure}
\caption{
Dependence on the pseudogap of physical quantities
associated with mean-field theory of a single chain.
The plot shows the coherence length $\xi_0$,
the width of the one-dimensional critical region $\Delta t_{1D}$,
and the inverse of the specific heat jump $\Delta C$.
All quantities are normalized to their rigid-lattice values.
For $\psi > 2.7 T$ the coefficient $b$ becomes
negative and the transition will be first order
(Section \protect\ref{secprop}).
The large reduction of $\Delta t_{1D}$ below the rigid lattice
value of 0.8 means that a mean-field treatment of the single
chain Ginzburg-Landau functional may be justified.
\label{figjump}}
\end{figure}

\begin{table}
\caption{
Comparison of experimental values for
K$_{0.3}$MoO$_3$ of various dimensionless ratios
with the predictions of two simple microscopic models.
The three-dimensional transition temperature is
$T_P= 183 $ K.
 The zero-temperature
energy gap $\Delta(0)$ is estimated from optical conductivity
data \protect\cite{deg}.
A Fermi velocity of $v_F=2 \times 10^5$ cm/sec was
estimated from band structure calculations \protect\cite{wha}.
$\Delta C$ is the specific heat jump
at the transition \protect\cite{bri} and $\gamma T_P$ is
the normal state electronic specific heat that has been
calculated from the density of states estimated from
magnetic susceptibility measurements \protect\cite{joh}
well above the transition temperature.
The longitudinal coherence length $\xi_{0z}$ has
been estimated from x-ray scattering experiments \protect\cite{gir}.
In both models the dimensionless ratios are independent
of any parameters,
except for $\Delta(0)/k_B T_P$ in
Schulz's model, which is described in Appendix \protect\ref{appsch}.
The rigid lattice theory \protect\cite{all,sch} involves a mean-field
treatment of the single-chain Ginzburg Landau
functional (\protect\ref{aa1}) with the
coefficients  (\protect\ref{mfa} - \protect\ref{mfc}).
 }
\begin{tabular}{lllcc}
Dimensionless & Experimental& Schulz & Rigid lattice\\*[-0.05in]
ratio & value &model & theory\\
\tableline
 $\displaystyle {\Delta(0) \over k_B T_P }$
 & $5 \pm 1$ & -- & 1.76\\
 $\displaystyle{{\Delta C \over \gamma T_P}}$
 & $5 \pm 1$ & 3.4 & 1.43\\
 $\displaystyle{{\xi_{z0} T_P \over v_F}}$
 & $0.18 \pm 0.04$ & 0.23 & 0.23 \\
\end{tabular}
\label{table1}
\end{table}

\begin{table}
\caption{Parameters for several quasi-one dimensional materials.
The observed transition temperature $T_P$ is always much smaller than
the rigid-lattice transition temperature
$T_{RL}$.  The phonons near
$2k_F$, which soften at the transition,
 can be treated classically since they have frequencies
of the order of $\Omega(0)$ (estimated from Raman and
neutron scattering) which is much smaller than $T_P$.
The zero-temperature gap $\Delta(0)$, estimated from
the peak in the optical absorption
was used to calculate  $T_{RL}$
 ($ T_{RL}=\Delta(0)/1.76k_B$).}
\begin{tabular}{lcccc}
& $T_P$ (K) & $\Delta(0)$ (meV)  & $T_P/T_{RL}$ & $\Omega(0)$ (K)\\
\tableline
K$_{0.3}$MoO$_3$&  183 & 90\tablenotemark[1] & 0.31 & 80
\tablenotemark[2] \\
(TaSe$_4$)$_2$I &  263 & 200$\tablenotemark[3]$ & 0.20 &
130\tablenotemark[4] \\
K$_2$Pt(CN)$_4$Br$_{0.3}$ & 120\tablenotemark[5]&  100\tablenotemark[6]
& 0.18 &  58 \tablenotemark[7]\\
TSeF-TCNQ        & 29   & 10\tablenotemark[8]  & 0.42 &  --  \\
\end{tabular}
\label{table2}
\tablenotetext[1]{ Ref. \cite{deg}}
\tablenotetext[2]{ J. P. Pouget, B. Hennion, C.
Escribe-Filippini, and M. Sato, Phys. Rev. B {\bf 43}, 8421 (1991)}
\tablenotetext[3]{ Ref. \cite{ber}}
\tablenotetext[4]{ S. Sugai, M.Sato, and S. Kurihara, Phys. Rev. B
{\bf 32}, 6809 (1985).}
\tablenotetext[5]{ Complete ordering does not occur \cite{car}.}
\tablenotetext[6]{Ref. \cite{bru}}
\tablenotetext[7]{Ref. \cite{car}}
\tablenotetext[8]{
 From activation energy of dc conductivity, Ref. \cite{sco}}

\end{table}

\onecolumn
\widetext
\appendix

\section{SCHULZ'S Model}
\label{appsch}

For completeness an alternative microscopic model
is discussed.
Schulz \cite{sch} considered only thermal fluctuations
in the phase of the order parameter.
He assumed that the temperature was sufficiently low that
fluctuations in the amplitude of the
order parameter were not significant.
(However, in real materials the amplitude fluctuations
{\it are} important \cite{mck}).
Fluctuations along the chain were treated exactly and
the interchain interactions were treated in the
mean-field approximation.
He derived a
free energy functional of the form (\ref{bg1}).
 The coefficients
in a tetragonal crystal ($a_x=a_y$) are
\begin{equation}
A={2 \over J} - {2 v_F \over \pi T^2}
\label{bg1a}
\end{equation}
\begin{equation}
B= {7 v_F^3 \over \pi^3 T^6}
\label{bg1b}
\end{equation}
\begin{equation}
C_x={a_x^2 \over 2 J}
\label{bg1c1}
\end{equation}
\begin{equation}
C_z={2 v_F^3 \over \pi^3 T^4}
\label{bg1c3}
\end{equation}

The coefficient $A(T)$ vanishes at the three-dimensional
transition temperature
\begin{equation}
 T_{3D} = \left({J v_F \over \pi}\right)^{1/2}.
\label{bbp1}
\end{equation}
Thus the ratio $\Delta(0)/k_B T_{3D}$ is not
a universal quantity.
The specific heat jump at the
transition is \begin{equation}
\Delta C= {1 \over a_x^2} {(A^\prime)^2 \over 2 B T_{3D}}
={16 \pi T_{3D} \over 7 a_x^2} = {24 \over 7} \gamma  T_{3D}
\label{bp1}
\end{equation}
where $\gamma T= 2\pi T/ 3 v_F a_x^2$ is the
normal state electronic specific heat.
The coherence length parallel to the chains is
\begin{equation}
\xi_{z0}= \left( {C_z  \over A^\prime}\right)^{1/2}
= {v_F   \over \sqrt{2} \pi T_{3D}}.
\label{br1}
\end{equation}
Equations (\ref{bp1}) and (\ref{br1}) then give the ratios
given in Table \ref{table1}.


\section{EVALUATION OF THE GINZBURG-LANDAU COEFFICIENTS}
\label{seccoeff}

The Ginzburg-Landau free energy functional (\ref{aa1})
is related to the partition function $Z$ by
the functional integral
\begin{equation}
{Z \over Z_0}= \int [d\phi(z)] \exp (-\beta F[\phi])
\label{part}
\end{equation}
where $Z_0$ is the partition function in the absence of interactions.

The phonon field in the Hamiltonian (\ref{hamel})
is treated classically.
For the Hamiltonian
\begin{equation}
H= \int dz \left\{ \Psi^\dagger (z) \bigg[ - iv_F \sigma_3 {\partial
\over \partial z} + {1 \over 2} \left(\Delta(z) \sigma_+
+ \Delta(z)^* \sigma_- \right) \bigg]\Psi (z)
+{ \Delta(z)^2 \over \lambda \pi v_F}\right\}
\end{equation}
the partition function is given by
\begin{eqnarray}
Z=&&\int d\Delta(z) \exp (- \beta\int dz { \Delta(z)^2 \over \lambda \pi v_F})
 \times \nonumber \\
&&\left\langle T \exp \left( -\int_0^\beta d \tau
\int dz \Psi^\dagger (z,\tau) \bigg[ - iv_F \sigma_3 {\partial
 \over \partial z} +
 {1 \over 2} \left(\Delta(z) \sigma_+ + \Delta(z)^* \sigma_- \right)
 \bigg]\Psi (z,\tau)\right)\right\rangle
\label{part2}
\end{eqnarray}
The goal is to get this expression into a form comparable to
(\ref{part}) so the coefficients $a,b,$ and $c$
can be extracted.

The linked cluster or cumulant expansion \cite{agd}
can be used to rewrite the time-ordered product in (\ref{part2}).
In general if the Hamiltonian is separated according to
\begin{equation}
H=H_0+H_1
\label{hamsep}
\end{equation}
and
\begin{equation}
\langle S \rangle \equiv \left\langle T \exp \left( - \int_0^{\beta}
d \tau H_1(\tau) \right) \right\rangle_0 = {Z \over Z_0}
\end{equation}
where $\langle .. \rangle_0 $ denotes a thermal average with respect to
$H_0$ then the linked cluster theorem states that
\begin{equation}
\langle S \rangle= \exp \left( \langle S \rangle_{0,{\rm conn}}\right)
-1
\end{equation}
where $\langle S \rangle_{0,{\rm conn}}$ denotes the set of connected
terms in the diagrammatic expansion of $S$.

The important question is how to make the separation (\ref{hamsep})?

A rigid-lattice treatment of the phonons neglects the effect of the
thermal lattice motion on the electronic states.
The Hamiltonian is separated according to
\begin{equation}
H_{0,RL}=
\int dz  \Psi^\dagger (z) \bigg[ - iv_F \sigma_3 {\partial
\over \partial z}  \bigg]\Psi (z)
\end{equation}
\begin{equation}
H_{1,RL}=  {1 \over 2} \int dz \Psi^\dagger (z)
\left(\Delta(z) \sigma_+ + \Delta(z)^* \sigma_- \right)
 \Psi (z)
\end{equation}
The resulting free-energy functional is
\begin{equation}
F_{RL}[\phi]=
\int dz \left\{{\phi(z)^2 \over \lambda \pi v_F}\right\}
+ {1 \over \beta} \left(\left\langle T \exp (- {1 \over 2}
\int dz \phi(z) \int_0^\beta d \tau
 \Psi^\dagger (z,\tau) \sigma_+ \Psi (z,\tau) -{\rm h.c.})
  \right\rangle_{{\rm o, RL, conn}} -1 \right)
  \label{fmf}
\end{equation}
where $\Delta(z)$ has been equated with $\phi(z)$ in the functional
integral. Expanding to fourth order in $\phi(z)$ gives the
rigid-lattice
coefficients given by (\ref{mfa} - \ref{mfc}) \cite{all}.

To improve on this rigid-lattice treatment we want to expand
relative to a Hamiltonian which includes at
least some of the effects of lattice fluctuations.
The rigid-lattice expression (\ref{fmf})
is modified  in the following way:
\begin{equation}
F[\phi]= \int dz \left\{{\phi(z)^2 \over \lambda \pi v_F}\right\}
+ {1 \over \beta} \left(\left\langle \left\langle T \exp (- {1 \over 2}\int dz
\phi(z)
\int_0^\beta d \tau
 \Psi^\dagger (z,\tau) \sigma_+ \Psi (z,\tau) -{\rm h.c.}) \right\rangle
 \right\rangle_{{\rm conn}} -1 \right)
 \label{free}
\end{equation}
where the expectation value of an operator $A$ is defined by
\begin{equation}
\langle \langle  A \rangle \rangle = \int { dv dv^* \over \pi}
 \exp (-v v^*) {\rm Tr} \big[ A \exp(- \beta H_0[v]) \big]
\end{equation}
with the Hamiltonian $H_0$
\begin{equation}
H_0[v]=\int dz \left\{ \Psi^\dagger (z) \bigg[ - iv_F \sigma_3 {\partial
 \over \partial z} + {1 \over 2} \psi ( v \sigma_+ + v^* \sigma_-)
 \bigg]\Psi (z)\right\}
\end{equation}
It is now possible to evaluate analytically the
free energy functional (\ref{free}).
The Fourier transform
\begin{equation}
\phi(z)={1 \over \sqrt{L}} \sum_q \phi_q e^{i q z}
\end{equation}
is performed
and the free energy (\ref{free}) expanded to fourth order in $\phi$.
The result is
\begin{equation}
F[\phi]=\sum_q a(q) \phi_q \phi_q^* + {1 \over {L}}\sum_{q_1,q_2,q_3}
b(q_1,q_2,q_3) \phi_{q_1}^* \phi_{q_2}^* \phi_{q_3} \phi_{q_1+q_2-q_3}
\end{equation}
where the coefficients $a(q)$ and $b(q_1,q_2,q_3)$ are given by
\begin{equation}
a(q)= {1 \over \pi v_F \lambda} + \pi T\sum_{\epsilon_n}\int{dvdv^*\over\pi}
e^{-vv*} \int{dk \over 2\pi}
Tr\left[\sigma_+\hat G(k+q,\epsilon_n,v)\sigma_-\hat
G(k,\epsilon_n,v)\right]\label{xfc}
\label{agreen}
\end{equation}
\begin{eqnarray}
b(q_1,q_2,q_3)= \pi T \sum_{\epsilon_n}\int{dvdv^*\over\pi}
e^{-vv*} \int{dk \over 2\pi}&&
Tr \ [\sigma_-\hat G(k+q_1,\epsilon_n,v)\sigma_-
\hat G(k+q_1+q_2,\epsilon_n,v) \nonumber \\
&&\sigma_+\hat G(k+q_1+q_2-q_3,\epsilon_n,v)\sigma_+
\hat G(k,\epsilon_n,v) ]\label{xfb}
\end{eqnarray}
and the electronic Green's function $\hat G(k,\epsilon_n,v)$
is defined in equation (\ref{ea2}).

Expanding $a(q)$ in powers of $q^2$ gives the Ginzburg-Landau coefficients
$a(T)$ and $c(T)$:
\begin{equation}
a(q)=a(T) + c(T) q^2 + ...
\end{equation}
The fourth-order coefficient is $b=b(0,0,0)$.
The coefficients $a,b,$ and $c$ in are given in terms of $\psi$
in equations (\ref{gla}-\ref{glc}).

\begin {references}
\bibitem{gin} V. L.Ginzburg, Fiz. Tverd. Tela
{\bf 2}, 2031 (1960) [Sov. Phys. Solid State {\bf 2}, 1824
 (1960)].
The width of the critical region, $\Delta T$, is defined by the
temperature at which the fluctuation contribution
to the specific heat below the transition temperature,
calculated in the Gaussian
approximation, equals the mean-field specific heat jump $\Delta C$.
It should be stressed that this gives only a very rough
estimate of the importance of fluctuations and that
there are several alternative definitions of the width
of the critical region.
Consequently, care should be taken when comparing estimates
from different references. This is particulary true
since different definitions can differ by
numerical  factors as large as $32 \pi^2$!
\bibitem{alternative}
There are different ways of handling the cutoff.
In evaluating this integral I have followed Rice
and Str\"assler \cite{ric} and Schulz \cite{sch}
and performed the integral over $q_\parallel$
without any cutoff while a cutoff is used for
$q_\perp$. A slightly different result is
obtained if a cutoff is included also for $q_\parallel$.
\end{references}
\end{document}